\documentclass[a4paper, 12pt]{article}
\usepackage{amsmath}
\begin{document}
\setlength{\baselineskip}{1.2\baselineskip}
\setlength{\parskip}{4ex}
\title{Rational coefficients of two-body energies in diagonal matrix elements of $j^n$ configurations}
\author{Igal Talmi\\Weizmann Institute of Science,\\Rehovot, Israel}
  \maketitle
  \begin{abstract}
Matrix elements of a two-body interaction between states of the
$j^n$ configuration (\textit{n} fermions in the \textit{j}-orbit)
are functions of two-body energies. In many cases, diagonal matrix
elements are linear combinations of two-body energies, whose
coefficients are rational and positive numbers. It is shown that if
in the $j^n$ configuration there is only one state with a given spin
\textit{J}, its eigenvalue (the diagonal matrix element) has
rational and positive coefficients of two-body energies. The
situation in cases with several \textit{J}-states in the $j^n$
configuration is more complicated and is discussed in detail. States
with this property are identified and constructed. In particular,
the relevance to states in the seniority scheme is shown.
\end{abstract}
\section{Introduction}
A system of \textit{n} identical fermions in a given
\textit{j}-orbit is considered. Such $j^n$ configurations of protons
or neutrons occur, for instance, in the shell model. Any
configuration contains a certain number of fully antisymmetric
independent states allowed by the Pauli principle. We consider shell
model Hamiltonians which contain mutual interactions between the
particles. The interactions considered are rotationally invariant
and hence, eigenstates of the Hamiltonian have definite values of
angular momentum \textit{J}. The mutual interaction is considered to
be a perturbation. Hence, only the sub-matrix of the Hamiltonian
defined by the states of the $j^n$ configuration is taken into
account. The states considered here are the states which define this
sub-matrix.

If there is only one state with given \textit{J} in the $j^n$
configuration, it is an eigenstate of the sub-matrix of the
Hamiltonian. Its eigenvalue is given by the corresponding diagonal
element of the sub-matrix. If there are several independent states
with the same value of \textit{J}, they define a smaller sub-matrix
whose diagonalization yields the eigenstates for a given mutual
interaction. If two-body mutual interactions are considered,
elements of the Hamiltonian sub-matrix of the $j^n$ configuration
are functions of two-body matrix elements. Due to the rotational
invariance of the interaction, the only non-vanishing two-body
matrix elements are between two-body states with the same value of
\textit{J}. Thus, these two-body matrix elements are the energies of
the two-body configuration.

In a recent paper [1], Zamick and Van Isacker consider partial
dynamical symmetries of certain states of the  (9/2)$^4$
configuration. They used their results to calculate also the
energies of the $J=2$ and $J=4$ states of the (7/2)$^4$
configuration. They express these energies as linear combinations of
two-body energies and in their arXiv: 0803.1569v1 (nucl-th) version
they "note that this derivation constitutes a proof that the
coefficients...in the energy expression...must be
    rational numbers".

This feature is quite general. There are many cases where diagonal
matrix element in $j^n$ configurations are linear combinations of
two-body energies whose coefficients are rational and positive
numbers. This feature has little relevance to numerical calculations
but it still has theoretical interest. The aim of this paper is to
study this property of states, to find the cases in which diagonal
elements are linear combinations of two-body energies whose
coefficients are positive rational numbers.

There is a simple way to calculate the diagonal matrix elements of a
two-body interaction in a system of \textit{n} fermions in a
\textit{j}-orbit. It is based on the use of \textit{m}-scheme wave
functions. These are determinantal wave functions characterized by a
set of quantum numbers $m_1,m_2,\ldots,m_n$, which are all
different. These wave functions are discussed in great detail in
ref.[2] where matrix elements of a two body interaction between such
states are derived. It was shown there that diagonal matrix elements
are equal to the \qquad \qquad \qquad \qquad $\sum<m_im_k|V|m_im_k>$
where the summation is over all pairs of \textit{m}-values in the
state considered.

Matrix elements of a rotationally invariant two-body interaction
between \textit{m}-states with given total $M=\sum m_i$ form a
sub-matrix. Its trace is equal to the sum of diagonal elements taken
between all states with given \textit{M}. States of the $j^n$
configuration with given values of \textit{J} and \textit{M}, $J\geq
M$, are linear combinations of these \textit{m}-~states. The linear
transformation, from \textit{m}-states to states with definite
\textit{J} and \textit{M}, is unitary or rather orthogonal. Hence,
the trace of the sub-matrix is invariant and may be used to
calculate diagonal matrix elements of a rotationally invariant,
\textit{M}-independent interaction in the \textit{J}-scheme.

Subtracting the trace of the $M=J-1$ sub-matrix from the one of
$J=M$, yields the sum  $\sum <j^n a J,M|V|j^n a J,M>$  where
\textit{a} characterizes the state if there are several states with
given \textit{J}  in the $j^n$ configuration. If there is only one
such state, this procedure yields its interaction energy. Otherwise,
only the sum of energies of all states with given \textit{J}  is
obtained. This case will be discussed in detail in the following.
The diagonal matrix elements in the \textit{m}-~scheme may be
expressed in terms of matrix elements of two particles coupled to a
given \textit{J} value. The two-particle matrix elements in the
\textit{m}-scheme may be transformed by using Clebsch-Gordan
coefficients as follows.
\begin{align}
    &<m_im_k|V|m_im_k>=\cr
    &\sum <j,m_i,j,m_k|j,j,J,M=m_i+m_k>^2<j^2JM|V|j^2JM>
\end{align}
Since the \textit{m}-states in (1) are antisymmetric, the sum on the
r.h.s. of (1) is only over \textit{even} values of \textit{J}
(states of two fermions with \textit{odd J}  values are symmetric).
Thus, the coefficients of the $V(J)= <j^2JM|V|j^2JM>$ in the
diagonal elements in the \textit{m}-scheme are equal to sums of
squares of Clebsch-Gordan coefficients. From formulae for these
coefficients (e.g. (15.36), ref.[2]) it is seen that their squares
are rational functions of their arguments. These arguments are the 3
\textit{J}-values and 3 \textit{m}-values.

Thus, the sum  $\sum_{\substack{J}}<j^n a J,M|V|j^n a J,M>$ is a
linear combination of two-body matrix elements
$V(J)=<j^2JM|V|j^2JM>$ which are independent of \textit{M}, whose
coefficients are rational and positive numbers. If there is only one
state with given \textit{J} in the $j^n$ configuration, it has the
"rational property" considered here. Its eigenvalue of a two-body
interaction is a linear combination of the \textit{V(J)} with
positive rational coefficients. If there are several such states, it
was shown above that this property holds only for the sum of their
diagonal elements. It will be shown in this paper which of these
states has this property. It will also be shown below, that in such
cases, not any arbitrary state may have this property.

States in the $j^n$ configuration with given \textit{J} must be
considered individually. This may be achieved by using coefficients
of fractional parentage (c.f.p.), introduced by Bacher and Goudsmit
[3] and intensively applied to atomic spectra by Racah[4]. An
antisymmetric wave function with \textit{n} particles and given
\textit{J} may be expanded as
\begin{align}
   \psi(j^n a J)=\Sigma [j^{n-1}(bJ_1)j J|\}j^n a J] \psi (j^{n-1}(b J_1)j_n a J)
\end{align}
\noindent In (2), \textit{a} and \textit{b} are used to specify
uniquely \textit{J} and $J_1$ states if there are several of them.
The wave functions on the r.h.s. of (2) are fully antisymmetric
functions of particles 1 to \textit{n}-1. They are usually not
antisymmetric in all \textit{n} particles but their linear
combination in (2) is.

\noindent Using the expansion (2), matrix elements in states of the
$j^n$ configuration may be expressed in terms of those in the
$j^{n-1}$ configuration. Due to the full antisymmetry of (2), it is
possible to calculate matrix elements of $\Sigma V_{ik}$ for $i,k<n$
whose number is (\textit{n}-1)(\textit{n}-2)/2 and multiply the
result by \qquad \qquad \qquad \qquad
\textit{n}(\textit{n}-1)/(\textit{n}-1)(\textit{n}-2)=\textit{n}/(\textit{n}-2)
to obtain the full value of the matrix element. This leads to
\begin{align}
 &<j^n a_n J| \Sigma V_{ik}|j^n b_n J>= \cr
 &n/(n-2)\sum_{a_{n-1}J_1} [j^{n-1}(a_{n-1}J_1)jJ|\}j^n a_nJ][j^{n-1}(b_{n-1}J_1)jJ|\}j^n
 b_nJ]
\times\cr &<j^{n-1}a_{n-1}J_1|\Sigma V_{ik}|j^{n-1} b_{n-1}J_1>
\end{align}
Using c.f.p. the matrix elements in (3) may be further reduced to
those in the $j^{n-2}$ configuration and so on until the $j^2$
configuration is reached. A direct result of (3) follows if there is
only one state with given \textit{J} in the configurations involved.
Then, diagonal matrix elements have the property considered here if
the squares of the c.f.p. are rational numbers. Since this was
proved above, the case with several states with the same value of
\textit{J} should be considered. This will be carried out in the
following, starting with the simple case of \textit{n=3} and
proceeding to higher \textit{n}  values.

\section{A simple case: the $j^3$ configuration}
A state with given \textit{J}  may be obtained by starting from a
\textit{principal parent} with given $J_0$ in the $j^2$
configuration, coupling to it $j_3$ and antisymmetrizing the
function $\psi(j^2(J_0)j_3J)$. If the resulting state is allowed,
the wave function does not vanish and its c.f.p. are given by (e.g.
(26.11), ref.[2] or (15.10), ref.[5])
\begin{align}
&[j^2(J_1)j J|\}j^3[J_0]J]=\cr &N(J_0,J)[\delta (J_1,J_0)+
2[(2J_0+1)(2J_1+1)]^{1/2}\{jjJ_1/JjJ_0\}]
\end{align}
where the normalization factor is given by
 \begin{align}
 N(J_0,J)=[3+6(2J_0+1)\{jjJ_0/JjJ_0\}]^{-1/2}
\end{align}
 and \textit{\{abc/def}\} are 6\textit{j}-symbols (Racah coefficients). The
c.f.p. of the principal parent is obtained by putting $J_1=J_0$  in
(4). Then from (5) follows the special $n=3$ case of the identity
proved in ref.[2],
\begin{align}
           n N(J_0,J)[j^2(J_0)j J|\}j^3[J_0]J]=1
\end{align}

To find the properties of the c.f.p., it is worth while to look at
the form of the 6\textit{j}-symbols. When Racah introduced these
symbols [6] (in a slightly different form and symmetry properties),
he also derived a closed formula of them in terms of their 6
arguments. In (15.38),ref.[2], it is expressed as
\begin{align}
\{{abc/def}\}=\Delta (abc) \Delta (aef) \Delta (dbf) \Delta
(dec)R(abcdef)
\end{align}
where
\begin{align}
           \Delta(abc)=[(a+b-c)!(b+c-a)!(c+a-b)!/(a+b+c+1)!]^{1/2}
\end{align}
and \textit{R} is a rational function of the 6 arguments. Hence, the
square of a 6\textit{j}-symbol is a rational function of its
arguments.  If a state obtained by antisymmetrization is not allowed
by the Pauli principle, the wave function vanishes.  This always
happens for states with \textit{J}=3\textit{j}-4 or
\textit{J}=${1\over2}$.  The unnormalized c.f.p. all vanish due to
the properties of the rational function in(7). From (7) and (8)
follow some simple properties of 6\textit{j}-symbols, like
\begin{align}
          \{jjJ_0/JjJ_0\}=\Delta(jjJ_0)^2\Delta (JjJ_0)^2 R(jjJ_0JjJ_0)
\end{align}
From (9) follows that the c.f.p. of the principal parent, with
$J_1=J_0$, is a square root of a rational function of \textit{j},
$J_0$ and \textit{J}. This is true also for the normalization
coefficient (5). Therefore, also the other c.f.p., with $J_1\neq
J_0$, share this feature since their squares are equal to
\begin{align}
          4N(J_0,J)^2(2J_0+1)(2J_1+1)\{j,j,J_1/J,j,J_0 \}^2
\end{align}
Thus, the squares of the c.f.p. of a state obtained by
antisymmetrization of a principal parent are rational functions of
the angular momenta. Hence, due to (3), such states have the
property considered here. Their eigenvalues of a two-body
interaction are linear combinations of the $V(J)= <j^2J|V|j^2J>$
with coefficients which are positive rational numbers. For
\textit{n}=3, the relation (3) yields directly these linear
combinations as follows
\begin{align}
 &<j^3 a J|\Sigma V_{ik}|j^3 a J>=\cr
 &3\Sigma [j^2(J_1)jJ|\}j^3aJ][j^2(J_1)jJ|\}j^3aJ] <j^2J_1|V|j^2J_1>
\end{align}
\par
If there is only one state with given \textit{J} in the $j^3$
configuration, the proof that it has the property considered, is
complete as already shown in the preceding section. In that case,
starting with all possible principal parents, those with $J_0$ which
can yield \textit{J}  when coupled with \textit{j}, the same state
is obtained after antisymmetrization (with a possible change of
overall sign). If there are several states with given \textit{J}  in
the $j^3$ configuration, they may be constructed by antisymmetrizing
\textit{several} principal parents. The diagonal matrix element of
each of them is given by the relation (11). Still, the states
obtained in this way need not be orthogonal. It will be shown now
that states obtained in the orthogonalization procedure have also
the property considered.

The orthogonalization of states with the same value of \textit{J}
constructed from principal parents is simplified by the following
property of such c.f.p. Due to (4), they may be expressed, for
$J_1\neq J_0$, as
 \begin{align}
 &[j^2(J_1)j J|\}j^3[J_0]J]=\cr
 &N(J_0,J)2[(2J_0+1)(2J_1+1)]^{1/2}\{jjJ_1/JjJ_0\}= \cr
    &N(J_0)(2J_0+1)^{1/2}\Delta (jjJ_0)\Delta (JjJ_0)(2J_1+1)^{1/2} \Delta(jjJ_1)\Delta
    (JjJ_1)R
\end{align}
where  $N(J_0)= N(J_0,J)$ (also in the following) and
$R=R(jjJ_1JjJ_0)$. Thus, these c.f.p. are products of a rational
function which multiplies a term which depends only on $J_0$ and
another one which depends only on $J_1$. The latter term is the same
for all states with various values of $J_0$.

To make a state $|b>$ orthogonal to a given state $|a>$, the latter
multiplied by $<a|b>$ should be subtracted from the former. The
orthogonalized state
\begin{align}
               |b^\prime >=|b>-<a|b>|a>
\end{align}
should then be normalized. Using c.f.p., the scalar product of the
two states is given by
\begin{align}
<j^n a J|j^n b
J>=\Sigma_{a_1J_1}[j^{n-1}(a_1J_1)jJ|\}j^naJ][j^{n-1}(a_1J_1)jJ|\}j^nbJ]
\end{align}
The state $|a\!>$ is taken to be constructed from the principal
parent $\psi(j^2(J_0)j_3J)$ and the $|b>$ state from
$\psi(j^2(J_0^\prime)j_3J)$. Making use of the c.f.p. (4) we find
that every term in the summation in (14), for $J_1\neq J_0, J_1\neq
J_0^\prime$ is given by
\begin{align}
    &N(J_0)N(J_0^\prime)2[(2J_0+1)(2J_1+1)]^{1/2}\{jjJ_1/JjJ_0\}\times\cr
    &2[(2J_0^\prime+1)(2J_1+1)]^{1/2}\{jjJ_1/JjJ_0^\prime\}
\end{align}
\par
From the properties (7) of 6\textit{j}-symbols follows that the
product of the two of them in (15) can be expressed as
\begin{align}
   &\{jjJ_1/JjJ_0\}\{jjJ_1/JjJ_0^\prime \}= \Delta (jjJ_1)\Delta (jjJ_0)\Delta (JjJ_1)\Delta
   (JjJ_0)\times\cr
&\Delta(jjJ_1)\Delta (jjJ^\prime_0) \Delta (JjJ_1)
\Delta(JjJ_0^\prime )R(jjJ_1JjJ_0)R(jjJ_1JjJ_0^\prime)=\cr
&\Delta(jjJ_0)\Delta (JjJ_0)\Delta (jjJ_0^\prime)\Delta
(JjJ_0^\prime)\Delta (jjJ_1)^2\Delta (JjJ_1)^2 \times\cr
&R(jjJ_1JjJ_0)R(jjJ_1JjJ_0^\prime)
\end{align}
Due to the form (7) of  6j-symbols, the r.h.s. of (16) may be
expressed as
\begin{align}
  &({jjJ_0/JjJ_0^\prime}/R(jjJ_0JjJ_0^\prime))\Delta(jjJ_1)^2
  \Delta(JjJ_1)^2\times\cr
  &R(jjJ_1JjJ_0)R(jjJ_1JjJ^\prime_0)
\end{align}
This expression is a 6\textit{j}-symbol multiplied by a rational
function, R($J_1$). Thus, every term in (14) becomes equal to
\begin{align}
  4N(J_0)N(J_0^\prime
  )[(2J_0+1)(2J_0^\prime+1)]^{1/2}\{jjJ_0/jjJ_0^\prime\}\times \cr&(2J_1+1)R(J_1)
\end{align}
The summation of terms with values of $J_1$ different from
$J_0^\prime$ and from $J_0$ may now be carried out and yields a
rational function multiplying
\begin{align}
       N(J_0)N(J_0^\prime )[(2J_0+1)(2J_0^\prime+1)]^{1/2}\{jjJ_0/jjJ_0^\prime \}
\end{align}
\par
The term in (14) with $J_1=J_0$ is equal to
\begin{align}
   &N(J_0)N(J_0^\prime
   )[1+2[(2J_0+1)(2J_0+1)]^{1/2}\{jjJ_0/JjJ_0\}]\times\cr
&2[(2J_0^\prime+1)(2J_0+1)]^{1/2}\{jjJ_0/JjJ_0^\prime \}
\end{align}
Due to the properties of 6\textit{j}-symbols (7), this expression is
equal to a rational function multiplying the expression (19). Also
the term with $J_1=J_0^\prime$ is equal to (19) multiplied by a
(different) rational function. Thus, the summation in (14) yields a
rational function multiplied by the factor (19).

To obtain the state which is orthogonal to the one characterized by
$J_0^\prime$ , eq.(13) is used. The unnormalized c.f.p. of that
state are defined by
\begin{align}
& N(J_0^\prime)[\delta (J_1,J_0^\prime )+
2[(2J_0^\prime+1)(2J_1+1)]^{1/2}\{jjJ_1/JjJ_0^\prime
 \}]-\cr
&N(J_0)N(J_0^\prime
)[(2J_0+1)(2J_0^\prime+1)]^{1/2}\{jjJ_0/jjJ_0^\prime \}\times\cr
&N(J_0)[\delta (J_1,J_0)+ 2[(2J_0+1)(2J_1+1)]^{1/2}\{jjJ_1/JjJ_0\}]R
\end{align}
where \textit{R} is a rational function. In the case that $J_1$ is
different from $J_0$ and $J_0^\prime$, (21) assumes the form
\begin{align}
      N(J_0^\prime)2[(2J_0^\prime+1)(2J_1+1)]^{1/2}\{jjJ_1/JjJ_0^\prime\}
      -\cr
N(J_0)N(J_0^\prime
)[(2J_0+1)(2J_0^\prime+1)]^{1/2}\{jjJ_0/jjJ_0^\prime \} \times \cr
N(J_0)2[(2J_0+1)(2J_1+1)]^{1/2}\{jjJ_1/JjJ_0\}R
\end{align}
In (22)  there are several square roots which appear twice. The
product of two 6\textit{j}-symbols is equal, according to (16) and
(17) to $\{jjJ_1/jjJ_0^\prime \}$ multiplied by a rational function.
Hence, the unnormalized c.f.p. of the orthogonalized state, given by
(22) are equal to the original c.f.p. multiplied by rational
functions.

In the case $J_1=J_0$, that c.f.p. is equal to
\begin{align}
   N(J_0^\prime)2[(2J_0^\prime+1)(2J_0+1)]^{1/2}\{jjJ_0/JjJ_0^\prime \}
   -\cr
N(J_0)N(J_0^\prime
)[(2J_0+1)(2J_0^\prime+1)]^{1/2}\{jjJ_0/jjJ_0^\prime \} \times\cr
N(J_0)[1+2[(2J_0+1)(2J_0+1)]^{1/2}\{jjJ_0/JjJ_0\}R
\end{align}
\noindent In (23),
$N(J_0)^2[1+2[(2J_0+1)(2J_0+1)]^{1/2}\{jjJ_0/JjJ_0\}]$ is a rational
function and hence, (23) is equal to the original c.f.p. multiplied
by a rational function, as are the other c.f.p. Also in the case
$J_1=J_0^\prime$ , the situation is similar. In that case, (21) is
equal to
\begin{align}
 N(J_0^\prime)[1+ 2[(2J_0^\prime+1)(2J_0^\prime+1)]^{1/2}\{jjJ_0^\prime/JjJ_0^\prime \}]
 -\cr
N(J_0)N(J_0^\prime
)[(2J_0+1)(2J_0^\prime+1)]^{1/2}\{jjJ_0/jjJ_0^\prime\} \times\cr
N(J_0)2[(2J_0+1)(2J_0^\prime+1)]^{1/2}\{jjJ_0^\prime/JjJ_0\}R
\end{align}
The c.f.p. of the principal parent of the original state is equal to
$N(J_0^\prime)$ multiplied by a rational function. The term
substracted from it is equal to $N(J_0^\prime)$ multiplied by the
product of the rational function \textit{R} and the square of a
product of square roots. Hence, all c.f.p. of the unnormalized
orthogonalized state are equal to the original ones multiplied by
various rational functions. The resulting state should be
normalized. It should be divided by the square root of the sum of
squares of these c.f.p. Since the squares of the latter are
rational, the normalization coefficient is the square root of a
rational function. Thus, the squares of the c.f.p. of the normalized
and orthogonalized state are rational and positive.

The c.f.p. of the orthogonalized and normalized states are equal,
apart from a common normalization factor, to the original c.f.p.
multiplied by rational functions. It may happen that certain c.f.p.
vanish after this procedure. For instance, starting from a certain
$J_0$, in the state with $J_0^\prime$ which was made orthogonal to
it, the c.f.p. for $J_1=J_0$ must vanish. If further
orthogonalizaions are necessary, due to existence of more
independent states with the same total \textit{J}, It is possible to
repeat the steps taken above.

Looking at the derivations above, it is clear that not all linear
combinations of states obtained from principal parents have the
property considered here. The c.f.p. may become sums of square roots
of relatively prime numbers and their squares will then not be
rational numbers. In fact, there are states whose diagonal matrix
elements are not linear combinations of two-body energies.  This may
happen to eigenstates obtained by diagonalization of the submatrix,
with given \textit{J}, of a two-body interaction. An example is
shown at the end of this section.  Naturally, it is possible to
construct linear combinations, other than those discussed above,
which have the "rational property".

Non-diagonal matrix elements have also a simple form when taken
between states sharing the property discussed here. We consider the
matrix element (3) for \textit{n=3} taken between different states.
We may take them to be the state obtained by antisymmetrizing  the
principal parent with $J_0$ and the one obtained from the
$J_0^\prime$ principal parent, orthogonalized as above. To calculate
the non-diagonal matrix element it is possible to make use of the
calculation of the overlap of the two states above. We start from
the formula (3) for the special case of \textit{n=3}
\begin{align}
&<j^3[J_0]J|\Sigma V_{ik}|j^3[J_0^\prime]J>=\cr
 &3\Sigma[j^2(J_1)j
J|\}j^3[J_0]J][j^2(J_1)jJ|\}j^3[J_0^\prime]J]<j^2J_1|V|j^2J_1>
\end{align}
Every term in the summation in (25) is equal to a corresponding
terms in(14) multiplied by $<j^2J_1|V|j^2J_1>$. Every term (14) is
transformed into (19), which is
\begin{align}
       N(J_0)N(J_0^\prime )[(2J_0+1)(2J_0^\prime+1)]^{1/2}\{jjJ_0/jjJ_0^\prime \}
\end{align}
multiplied by a rational function. Hence, the coefficient of
$<j^2J_1|V|j^2J_1>$  in (25) is (26) multiplied by a rational
function. Thus, the non-diagonal matrix elements in the scheme of
states described above, are linear combinations of two-body energies
with rational functions multiplied by a common square root of a
rational function.

Before proceeding to higher values of \textit{n}, it it worth-while
to look at  some simple examples in $j^3$ configurations. States
with \textit{J=j} (for \textit{n} odd) and seniorities \textit{v=1}
of the $j^n$ configuration are obtained from principal parents with
$J_0=0$ and $v_0=0$ in the $j^{n-1}$ configuration. The seniority
scheme is explained in detail in refs.[2] and [5]. It is also
reviewed in the last section of this paper.  In the case of
\textit{n=3}, putting $J_0=0$ in (4) the following results were
obtained
\begin{align}
 &[j^2(0)j J=j|\}j^3 v=1J=j]=\{1+2\{jj0/jj0\}N(0)=\cr
 &[(2j-1)/3(2j+1)]^{1/2}\cr
&[j^2(J_1)j J=j|\}j^3 v=1 J=j]=\cr
&2(2J_1+1)^{1/2}\{jjJ_1/jj0\}N(0)=-2[(2J_1+1)/3(2j-1)(2j+1)]^{1/2}
\end{align}
As long as $j<9/2$, there is only one state with \textit{J=j} in the
$j^3$ configuration and that state has c.f.p. given above. For
higher values of \textit{j} there are other states with \textit{J=j}
and they all have seniorities \textit{v=3}. Their number is
$[(2j-3)/6]$ where $[x]$ is the largest integer not exceeding $x$.
They may be constructed from principal parents with even values of
$J_0>0$ and then orthogonalized to the \textit{v=1} state and to
each other. The first part of this procedure, for any \textit{j}, is
carried out in ref.[5], Section 20, pages 396 - 398.

The two states with \textit{J}=9/2 in the $j^3$ configuration offer
an opportunity to demonstrate a statement made above. Not all states
have the property considered in this paper. There are states whose
diagonal matrix elements are not equal to linear combinations of
two-body energies. It was mentioned that such cases occur in
diagonalization of the Hamiltonian sub-matrix defined in the $j^n$
configuration by the states with a given value of \textit{J}.

Consider the 2x2 sub-matrix defined by the \textit{v}=1 and
\textit{v}=3 states with \textit{J}=9/2. Both diagonal matrix
elements $V_1$ and $V_3$ are linear combinations of the
\textit{V(J)} with positive rational coefficients. The non-diagonal
matrix element \textit{V} is also a linear combination of the
\textit{V(J)} with rational coefficients multiplied by the square
root of a rational function. The two eigenvalues with \textit{J}=9/2
are the roots of the quadratic secular equation
\begin{align}
  x^2-(V_1+V_3)x+V_1V_3-V^2=0
\end{align}
They are equal to
\begin{align}
 & x={1\over 2}\{V_1+V_{3}\pm [(V_1+V_3)^2-4V_1V_3+4V^2]^{1/2}\}=\cr
&{1\over 2}\{V_1+V_{3}\pm [(V_1-V_3)^2+4V^2]^{1/2}\}
\end{align}
Both eigenvalues are sums of linear combinations with rational
coefficients and square roots of quadratic functions of the
\textit{V(J)}.

In some special cases, the square root may turn out to be a linear
function of the \textit{V(J)}. In the case of the $J=9/2$ described
above, this cannot happen. The $v=3$ state is orthogonal to the
$v=1$ state whose principal parent is the $j^2(J_1=0)$ state. Hence,
the c.f.p. with $J_1$=0 of the \textit{v}=3 state vanishes. Thus, in
the linear combination $V_3$ the coefficient of \textit{V}(0)
vanishes. This coefficient vanishes also in the non-diagonal matrix
element \textit{V}. The term $V_1-V_3$ may be expressed as a linear
combination of \textit{V}(0) and terms with $V(J), J>0$ even,
$c_0V(0)+\Sigma c_JV(J)$. Its square is equal to
$c_0^2V(0)^2+2c_0V(0)\Sigma c_JV(J)+(\Sigma c _JV(J))^2$. The term
with $V^2$ has the form $(\Sigma d_JV(J))^2$ with no \textit{V}(0)
terms, and adding it cannot change the quadratic and linear terms in
\textit{V}(0). If \textit{V} does not vanish, adding 4$V^2$ to
$(V_1-V_3)^2$ cannot yield a square of a rational function of the
\textit{V}(\textit{J}). Hence, in the example considered here, the
eigenvalues of a two-body interaction are definitely not linear
combinations of the two-body energies.

There is no contradiction between this result and (11). There, the
diagonal matrix element of a state is a linear combination of the
\textit{V}(\textit{J}) with coefficients which are squares of the
c.f.p.(multiplied by 3). The states constructed in (4), as well as
all states constructed in this paper, are states which form a basis
of all states with given \textit{J}. Any antisymmetric state may be
expanded in terms of c.f.p. as in (2). The states constructed in
(4), as well as others, are based on the use of principal parents.
Hence, the c.f.p. are functions of the angular momenta and are
independent of the two-body interaction. If there is only one state
with given \textit{J}, it is an eigenstate and its eigenvalue is a
linear combination of the \textit{V}(\textit{J}). If there are
several such states and the Hamiltonian sub-matix has non-vanishing
non-diagonal elements, this is no longer the case. Eigenstates may
be expanded in terms of c.f.p. but the latter may well be
(irrational) functions of the two-body matrix elements. This is
indeed the case in the example shown above.

\section{General $j^n$ configurations}
\par
In the preceding section it was shown that for the \textit{n=3}
case, states constructed by antisymmetrizing principal parents have
the property considered in this paper. Diagonal matrix elements of a
two-body interaction are linear combinations of two-body matrix
elements (energies) \textit{V(J)} whose coefficients are rational
numbers. This property is still shared by such states which were
orthogonalized to each other. These features follow from
demonstrating that for \textit{n=3}, the c.f.p. of these states are
square roots of rational functions of the angular momenta involved.
It was shown that the normalization factor is a square root of a
rational function. That the c.f.p. of the principal parent is equal
to the normalization factor multiplied by a rational function. There
is a general relation between this c.f.p. and the normalization
factor. It was proved in ref.[2], (26.23), that
\begin{align}
         n[j^{n-1}(a,J_0)jJ|\}j^n[a,J_0]J] N(J_0,J)=1
\end{align}
In the following it will be shown by induction that these features
are shared also by c.f.p. of states in $j^n$ configurations for
$n>3$. Starting from \textit{n=3} it will be shown that if these
features hold for c.f.p. in the $j^{n-1}$ configuration, they hold
also for c.f.p. in the $j^n$ configuration. The property considered
in this paper, that the expectation value of a two-body interaction
is then a linear combination of two-body energies whose coefficients
are rational functions then follows as will be shown below.

Consider a state of the $j^n$ configuration obtained by
antisymmetrizing a principal parent  $\psi (j^{n-1}(a J_0)j_n J)$.
If the result does not vanish, its c.f.p. can be expressed by c.f.p.
of states in the $j^{n-1}$ configuration. All states considered in
the following are taken to be allowed states whose wave functions do
not vanish. This relation (ref.[2] eq.(26.30) or (15.29) of ref.[5])
is
\begin{align}
   &n[
   j^{n-1}(a_{n-1},_0J_{n-1},_0)jJ_n|\}j^n[a_{n-1},_0J_{n-1},_0]J_n]\times\cr
&[j^{n-1}(a_{n-1},_1J_{n-1},_1)jJ_n|\}j^n[a_{n-1},_0J_{n-1},_0]J_n]=\cr
 &\delta( a_{n-1},_1,a_{n-1},_0) \delta(J_{n-1},_1,J_{n-1},_0)+\cr
 &(n-1)
\Sigma(-1)^{J_{n-1,0}}+J_{n-1,1}[(2J_{n-1,0}+1)(2J_{n-1,1}+1)]^{1/2}\cr
&\{J_{n-2}jJ_{n-1,1}/J_njJ_{n-1,0}\}[j^{n-2}(a_{n-2}J_{n-2})jJ_{n-1,0}|\}j^{n-1}a_{n-1,0}J_{n-1,0}]\times\cr
&[j^{n-2}(a_{n-2}J_{n-2})jJ_{n-1,1}|\}j^{n-1}a_{n-1,1}J_{n-1,1}]
\end{align}
The first step in calculating the c.f.p. is to put in (31),
$a_{n-1,1}=a_{n-1,0}$  and $J_{n-1,1}=J_{n-1,0}$. This leads to the
result
\begin{align}
    &n[J^{n-1}(a_{n-1,0}J_{n-1,0})jJ_n| \}j^n[a_{n-1,0}J_{n-1,0}]J_n]^2
    =\cr
&1+(n-1)(-1)^{2J_0}\Sigma (2J_{0+1})\{J_2jJ_0/JjJ_0\}\times\cr
&[j^{n-2}(a_{n-2}J_{n-2})jJ_{n-1,0}|\}j^{n-1}a_{n-1,0}J_{n-1,0}]^2
\end{align}
As in (9), the 6 \textit{j}-symbol in (32) is a rational function of
its arguments. If the squares of the n-2 $\rightarrow $  n-1 c.f.p.
are rational, as assumed for the induction, then the c.f.p. on the
l.h.s. of (32) is indeed the square root of a rational function for
any \textit{n}. According to (30), the normalization coefficient is
also a square root of a rational function.

The c.f.p. of all states are given by the r.h.s. of (31) divided by
\textit{n} multiplying the c.f.p. of the principal parent. Due to
(30), this amounts to multiplying the r.h.s. of (31) by the
normalization coefficient $N(J_{n-1,0}J_n)$. Starting from the known
c.f.p. for the \textit{n}=3 case, it seems  that in general the
c.f.p.
$[j^{n-1}(a_{n-1,1}J_{n-1,1})jJ_n|\}j^n[a_{n-1,0}J_{n-1,0}]J_n]$ is
equal to a rational function of the various angular momenta
multiplied by
\begin{align}
   N(J_{n-1,0}J_n)\Pi N(J_{r0}J_{r+1,0})\Pi
   N(J_{r1}J_{r+1,1})\Pi [(2J_{r0}+1)(2J_{r1}+1)]^{1/2}\times\cr
\Pi \Delta (J_{r0}jJ_{r+1,0}) \Pi\Delta (J_{r1}jJ_{r+1,1}) \Delta
(J_{n-1,0} j J_n)\Delta (J_{n-1,1} j J_n)
\end{align}
\noindent
In the products in (33), $J_{1,0}=J_{1,1}=j$ and
\textit{r} goes up to \textit{n}-2. In the following, this form of
c.f.p. will be proved by induction. Apart from the first factor,
(33) is a product of two expressions which are the same functions of
the various $J_{r0}$ and the various $J_{r1}$. The former define the
state considered and the latter, the component whose c.f.p. is
selected. This feature was noticed in the \textit{n}=3 case and is
in agreement with the property of c.f.p. given by (26.21) of ref[2].
The various angular momenta belong to wave functions which were
obtained by antisymmetrization of principal parents and by
orthogonalization if necessary. The intermediate angular momenta
supply the additional quantum numbers, denoted above by $a_{r0}$ and
$a_{r1}$, needed to specify the states considered. This will be the
case for all states considered below.

It is worth-while to point out that the form (33) is the correct one
also for the c.f.p. of the principal parent. Putting $J_{r1}=J_{r0}$
for all \textit{r}, the result is a rational function multiplying
the normalization factor
\begin{align*}
               N(J_{n-1,0}J_n)=n N(J_{n-1,0}J_n)^2/n
               N(J_{n-1,0}J_n)\cr
\end{align*}
According to (30) this is equal to a rational function multiplying
the c.f.p. of the principal parent. Hence, no special care should be
paid to that c.f.p. The fact that (33) is the correct expression of
all c.f.p. will be proved by induction with respect to \textit{n}.
Before doing it, let us draw some conclusions from the structure
(33) of c.f.p.

The first conclusion which follows from (33) is that the squares of
these c.f.p. are positive rational functions of the various angular
momenta. This was shown above for the c.f.p. of the principal parent
and this feature is shared by all of them. The aim of this paper is
to show for which states the expectation values of a two-body
interaction are linear combinations of two-body energies with
rational coefficients. When (33) will be shown to be correct, the
answer to this question will be given. States with this property are
those obtained by consecutive constructions from principal parents
and antisymmetrizations.

Another property of c.f.p. which follows from (33) is important for
the construction of orthogonal sets of states with the same total
spin \textit{J}. As in the case of \textit{n}=3, we construct such a
set by choosing one state due to a principal parent. If there are
other such independent states, obtained from other principal
parents, they are made orthogonal to it. The orthogonalization is
carried out by using (13). First, the scalar product (14) is
calculated, for the state (33) defined by $J^\prime_{r0}$ and the
one defined by $J_{r0}$. It is a linear combination and each of its
terms is a rational function multiplied by
\begin{align}
    &N(J_{n-1},_0J_n)\Pi N(J_{r0}J_{r+1,0})\Pi
    N(J_{r1}J_{r+1,1})\Pi[(2J_{r0}+1)(2J_{r1}+1)]^{1/2}\times\cr
&\Pi \Delta(J_{r0}jJ_{r+1,0}) \Pi \Delta (J_{r1}jJ_{r+1,1})\Delta
(J_{n-1,0} j J_n)\Delta (J_{n-1,1} j J_n)\times \cr
&N(J_{n-1,0}^\prime J_n)\Pi N(J_{r0}^\prime J_{r+1,0}^\prime)\Pi
N(J_{r1}J_{r+1,1})\Pi[(2J_{r0}^\prime+1)(2J_{r1}+1)]^{1/2}\times \cr
&\Pi\Delta (J_{r0}^\prime jJ_{r+1,0}^\prime )\Pi\Delta (J_{r1}j
J_{r+1,1}) \Delta(J_{n-1,0}^\prime jJ_n) \Delta(J_{n-1,1} j J_n)
\end{align}
In (34) all terms with $J_{r1}$ appear twice and thus contribute
only to the rational function multiplying it. Hence, the summation
over $J_{n-1,1}$ in (14) may be carried out yielding for the scalar
product a rational function multiplying
\begin{align}
    &N(J_{n-1,0}J_n) N(J_{n-1,0}^\prime J_n)\Pi N(J_{r0}J_{r+1,0}) \Pi
    N(J_{r0}^\prime J_{r+1,0}^\prime)\times \cr
&\Pi[(2J_{r0}^\prime+1)(2J_{r0}+1)]^{1/2}\Pi\Delta
(J_{r0}jJ_{r+1,0}) \Pi\Delta (J_{r0}^\prime jJ_{r+1,0}^\prime)\times
\cr &\Delta(J_{n-1,0} j J_n) (J_{n-1,0}^\prime jJ_n)
\end{align}
\par
According to (13), the state with  $J_{r0}$ should be multiplied by
the scalar product (35) and subtracted from the state with
$J_{r0}^\prime$ which is to be orthogonalized to the other state.
The factor (35) multiplied by the rational function is independent
of $J_{n-1,1}$. Multiplying it by a c.f.p. of the state with
$J_{r0}$, we obtain according to (35) and (33), a rational function
multiplying
\begin{align}
 &N(J_{n-1,0}J_n) N(J_{n-1,0}^\prime J_n)\Pi N(J_{r0}J_{r+1,0}) \Pi N(J_{r0}^\prime
  J_{r+1,0}^\prime)\times \cr
&\Pi[(2J_{r0}^\prime+1)(2J_{r0}+1)]^{1/2}\Pi\Delta
(J_{r0}jJ_{r+1,0}) \Pi\Delta (J_{r0}^\prime jJ_{r+1,0}^\prime
)\times \cr &\Delta(J_{n-1,0} j J_n)\Delta (J_{n-1,0}^\prime
jJ_n)\times \cr &N(J_{n-1,0}J_n)\Pi N(J_{r0}J_{r+1,0})\Pi
N(J_{r1}J_{r+1,1})\Pi[(2J_{r0}+1)(2J_{r1}+1)]^{1/2}\times \cr
&\Pi\Delta (J_{r0}jJ_{r+1,0}) \Pi \Delta (J_{r1}jJ_{r+1,1})\Delta
(J_{n-1,0} j J_n)\Delta (J_{n-1,1} j J_n)
\end{align}
In (36), all terms with the various  $J_{r0}$  are squared and may
be absorbed in the rational function which multiplies the following
expression
\begin{align}
           &N(J_{n-1,0}^\prime J_n)\Pi N(J_{r0}^\prime J_{r+1,0}^\prime )\Pi
           N(J_{r1}J_{r+1,1})\times \cr
&\Pi[(2J_{r0}^\prime+1)(2J_{r1}+1)]^{1/2} \Pi\Delta (J_{r0}^\prime
jJ_{r+1,0}^\prime )\Pi\Delta (J_{r1}jJ_{r+1,1})\times \cr
&\Delta(J_{n-1,0}^\prime jJ_n) \Delta(J_{n-1,1} j J_n)
\end{align}
\par
The expression (37) is identical with the corresponding c.f.p. of
the state with $J_{r0}^\prime$ multiplied by a rational function.
Hence, also the c.f.p. of the orthogonalized state are square roots
of rational functions and have the form given by (33). All states
with the same value of $J_n$ have c.f.p. given by (33). They are
different by a common factor to all c.f.p. of each state which is a
square root of a rational function. They are orthogonal to each
other due to the different rational functions multiplying the
expression (33) to obtain the full c.f.p. Some of these rational
functions may vanish. The states orthogonalized in this way are not
normalized and must be explicitely normalized.

We turn now to the form (33) of c.f.p. of states constructed
consecutively from principal parents and antisymmetrized. Obviously,
it holds for the case \textit{n=3}. To demonstrate that it is
correct also for c.f.p. with $n>3$ we use the recursion relation
(31) for induction by \textit{n}. We show that if (31) is the
correct form for a given \textit{n-1}, it holds also for \textit{n}.
According to the assumption, every term in the summation in (31) is
given by a rational function multiplying the expression
\begin{align}
   &N(J_{n-2,0}J_{n-1,0})\Pi N(J_{r0}J_{r+1,0})\Pi
   N(J_{r1}J_{r+1,1})\Pi[(2J_{r0}+1)(2J_{r1}+1)]^{1/2}\times \cr
&N(J^\prime _{n-2,0} J^\prime _{n-1,0})\Pi N(J_{r0}^\prime
J_{r+1,0}^\prime)\Pi
N(J_{r1}J_{r+1,1})\Pi[(2J_{r0}^\prime+1)(2J_{r1}+1)]^{1/2} \times
\cr &\Pi\Delta (J_{r0}jJ_{r+1,0}) \Pi \Delta(J_{r1}jJ_{r+1,1})\Delta
(J_{n-2,0} j J_{n-1,0}) \Delta(J_{n-2,1} j J_{n-1,0})\times \cr
&\Pi\Delta (J_{r0}^\prime jJ_{r+1,0}^\prime)\Pi\Delta (J_{r1}jJ
_{r+1,1}) \Delta(J_{n-2,0}^\prime j J_{n-1,0}^\prime)
\Delta(J_{n-2,1} j J_{n-1,0}^\prime)\times \cr
&\{J_{n-2,1}jJ_{n-1,0}^\prime/J_njJ_{n-1,0}\}
\end{align}
\par
 This expression may be simplified by noting that there are
several identical factors which may be absorbed in the rational
function. Thus, we obtain instead of (38) the form
\begin{align}
          &\Pi N(J_{r1}J_{r+1,1})^2\Pi(2J_{r1}+1) \Pi
          \Delta(J_{r1}jJ_{r+1,1})^2 \times \cr
&\Pi N(J_{r0}J_{r+1,0})\Pi N(J_{r0}^\prime J^\prime_{r+1,0})
\Pi[(2J_{r0}+1)(2J_{r0}^\prime+1)]^{1/2}\times \cr &\Pi\Delta
(J_{r0}jJ_{r+1,0}) \Pi \Delta(J_{r0}^\prime
jJ_{r+1,0}^\prime)\{J_{n-2,1}jJ_{n-1,0}^\prime/J_njJ_{n-1,0}\}\times
\cr &\Delta(J_{n-2,1} j J_{n-1,0})\Delta (J_{n-2,1} j
J_{n-1,0}^\prime)
\end{align}
\par
Whereas in the products in (38) \textit{r} goes up to \textit{n}-3,
in (39) it goes up to \textit{n}-2. Using the formula (7) for the
6\textit{j}-symbol, the product
\begin{align}
      \Delta(J_{n-2,1}j J_{n-1,0}^\prime)\Delta (J_{n-2,1} j J_{n-1,0})\{J_{n-2,1}jJ_{n-1,0}^\prime/J_njJ_{n-1,0}\}
\end{align}
in (39) becomes equal to a rational function multiplied by
\begin{align}
    \Delta(J_{n-2,1}j J_{n-1,0})^2\Delta (J_{n-2,1} j J_{n-1,0}^\prime)^2\Delta (J_{n-1,0}j J_n) \Delta (J_{n-1,0}^\prime j J_n)
\end{align}
Thus the $J_{n-2,1}$ angular momentum appears as an argument only of
rational functions and the summation over it can be directly carried
out for all terms. The result of  this summation is a rational
function multiplied by
\begin{align}
       \Pi N(J_{r0}J_{r+1,0})\Pi N(J_{r0}^\prime J_{r+1,0}^\prime
       )\Pi[(2J_{r0}+1)(2J_{r0}^\prime +1)]^{1/2}\times \cr
\Pi\Delta (J_{r0}jJ_{r+1,0}) \Pi\Delta (J_{r0}^\prime
jJ_{r+1,0}^\prime)\Delta (J_{n-1,0}j J_n)\Delta (J_{n-1,0}^\prime j
J_n)
\end{align}
According to (31), this expression should be multiplied by
$N(J_{n-1,0}J_n)$ and is then identical with (33).

Thus, if the form (33) is correct for \textit{n}-1 it is correct
also for \textit{n}. Since it is correct for \textit{n=3}, it holds
for any value of \textit{n}. From (33) follows that the squares of
c.f.p. obtained from principal parents and antisymmetrization are
rational functions. According to (3), this is sufficient to
guarantee that if in every $j^r$ configuration, $r\leq n$, there is
only one state with a given value of $J_r$, the diagonal matrix
elements have the property considered in this paper. This special
case was discussed in the Introduction in a simpler way. Relation
(3) becomes more complicated if there are several states with the
same value of \textit{J}. Unlike the situation in the simple
\textit{n=3} case, using relation (3) for diagonal matrix elements
in the $j^n$ configuration may well lead to non-diagonal matrix
elements in $j^r$ configurations with $r\leq n$. Non-diagonal matrix
elements are not given by squares of $n-1\rightarrow n$ c.f.p. There
is, however, a way to make use of the structure (33) of these c.f.p.
to show that the states constructed in this way have the property
considered in this paper.

Any antisymmetric wave function in the $j^n$ configuration may be
expanded in $n-2\rightarrow n$ c.f.p. which are discussed in detail
in ref.[2]. They are defined by
\begin{align}
  \psi(j^naJ)= \Sigma[j^{n-2}(bJ_1)j^2(J^\prime )J|\}j^naJ]\psi (j^{n-2}(bJ_1)j_{n,n-1}^2 (J^\prime )aJ)
  \end{align}
Such an expansion is always possible since the wave functions on the
r.h.s. of (43) form a complete basis for the space of $j^n$ states
with given \textit{J}, which are antisymmetric in particles 1 to
\textit{n-2} and in particles \textit{n,n-1}. Fully antisymmetric
$j^n$ wave functions with given \textit{J} are included in this
space and hence, may be expanded as in (43). The calculation of
two-body matrix elements of a scalar operator, like the mutual
interaction between particles, is very simple if such c.f.p. are
used. Since these states are fully antisymmetric, it is possible to
calculate matrix elements of one term in the interaction,
$V_{n,n-1}$ say, and multiply the result by the number of such
terms, $n(n$-1)/2. Thus, using the expansion (43), we obtain the
result
\begin{align}
     &<j^naJ|\Sigma V_{ik}|j^naJ>=n(n-1)<j^naJ|V_{n,n-1}|j^naJ>/2=\cr
&[n(n-1)/2]\Sigma [j^{n-2}(bJ_1)j^2(J^\prime )J|\}j^naJ]
[j^{n-2}(b"J_1")j^2(J ")J|\}j^naJ]\times \cr
        &<j^{n-2}(bJ_1)j^2(J^\prime )J|V_{n,n-1}|j^{n-2}(b"J_1" )j^2(J")J>
        \end{align}
        \par
Integration of (44) over coordinates 1 to \textit{n}-2 yields zero
unless $J_1=J_1"$ and $b"\equiv b$. Since $V_{n,n-1}$ is a scalar
operator, its matrix element on the r.h.s. of (44) vanishes if $J"$
is not equal to $J^\prime$. If they are equal, the two-body matrix
element in (44) is equal to $<j^2J^\prime|V |j^2J^\prime
>$. Hence, the matrix element is given by
\begin{align}
&<j^n a J|\Sigma V_{ik}|j^n a J >=\cr
 &[n(n-1)/2]\Sigma_{b J_1 J^1} [j^{n-2}(bJ_1)j^2(J^\prime )J|\}j^naJ]^2
 <j^2J^\prime|Vj^2J^\prime >
 \end{align}
Hence, using the $n-2\rightarrow n$ c.f.p. in(43), diagonal matrix
elements of a two-body interaction are given by the linear
combination (45). In it, the coefficients of the two-body energies
are squares of the c.f.p. multiplied by $n(n$-1)/2. The expectation
value in the state $|j^naJ>$  of a two-body interaction is a linear
combination of two-body  energies with coefficients that are
positive rational functions if the squares of these c.f.p. are
rational.

To find out which states have $n-2\rightarrow n$ c.f.p. whose
squares are rational functions, it is convenient to express them in
terms of $n-1\rightarrow n$ c.f.p. This expression is given by
(26.44) in ref.[2] as
\begin{align}
   &[j^{n-2}(a_{n-2,0}J_{n-2,0})j^2(J^\prime )J_n|\}j^n a J_n]=\cr
&\Sigma[j^{n-2}(a_{n-2,0}J_{n-2,0})jJ_{n-1,1}|\}j^{n-1}a_{n-1,1}J_{n-1,1}]\times
\cr &[J^{n-1}(a_{n-1,1}J_{n-1,1})jJ_n|\}j^{n}aJ_n]\times \cr
&[(2J_{n-1,1}+1)(2J ^\prime +1)]^{1/2}
\{J_{n-2,0}jJ_{n-1,1}/jJ_nJ^\prime \}
\end{align}
The form (33) was proved above to be valid for c.f.p. of certain
states constructed from principal parents. Using such states in
(46), the form of their $n-1\rightarrow n$ c.f.p. may be taken from
(33). Thus, the $n-2\rightarrow n-1$ c.f.p.
$[j^{n-2}(a_{n-2,0}J_{n-2,0})jJ_{n-1,1}|\}j^{n-1}[a_{n-2,1},J_{n-2,1}]
J_{n-1,1}]$ is expressed as a rational function multiplying
\begin{align}
 &N(J_{n-2,1}J_{n-1,1})\Pi N(J_{r1}J_{r+1,1})\Pi
 N(J_{r0}J_{r+1,0})\Pi[(2J_{r1}+1)(2J_{r0}+1)]^{1/2}\times \cr
&\Pi\Delta (J_{r1}jJ_{r+1,1})\Pi\Delta (J_{r0}jJ_{r+1,0})\Delta
(J_{n-2,1} j J_{n-1,1})\Delta (J_{n-2,0} j J_{n-1,1})
\end{align}
The $n-1\rightarrow n$ c.f.p. ,$[j^{n-1}([a_{n-2,1},J_{n-2,1}]
J_{n-1,1})jJ_n|\}j^{n}[J_{n-1,0}^\prime]J_n]$ in (46) can be
similarly expressed, as a rational function multiplying the
following expression which is written in more detail than in (33)
\begin{align}
  &N(J_{n-1,0}^\prime J_n)\Pi N(J_{r0}^\prime J_{r+1,0}^\prime)\Pi
  N(J_{r1}J_{r+1,1})N(J_{n-2,1}J_{n-1,1})\times \cr
& \Pi[(2J_{r0}^\prime+1)(2J_{r1}+1)]^{1/2}
[(2J_{n-1,0}^\prime+1)(2J_{n-1,1}+1)]^{1/2}\times \cr &\Pi\Delta
(J_{r0}^\prime jJ_{r+1,0}^\prime ) \Pi\Delta
(J_{r1}jJ_{r+1,1})\Delta (J_{n-2,1}jJ_{n-1,1})\times \cr
&\Delta(J_{n-1,0}^\prime j J_n)\Delta (J_{n-1,1} j J_n)
\end{align}
\par
Substituting these expressions into (46) yields for each term in the
summation a rational function multiplied by
\begin{align}
     &N(J_{n-1,0}^\prime J_n) N(J_{n-2,1}J_{n-1,1})^2\times
\Pi N(J_{r1}J_{r+1,1})^2\Pi N(J_{r0}J_{r+1,0})\cr &\Pi
N(J_{r0}^\prime J_{r+1,0}^\prime) \Pi[(2J_{r0}+1)(2J_{r0}^\prime
+1)]^{1/2} \Pi(2J_{r1}+1)\times\cr &[(2J_{n-1,0}^\prime +1)(2J
^\prime +1)]^{1/2} \Pi \Delta(J_{r0}jJ_{r+1,0}\Pi
\Delta(J_{r0}^\prime jJ_{r+1,0}^\prime )\times\cr\ &\Pi\Delta
(J_{r1}jJ_{r+1,1})^2 \Delta(J_{n-2,0} j J_{n-1,1})
\Delta(J_{n-2,1}jJ_{n-1,1})^2 \Pi \Delta(J_{r0}jJ_{r+1,0})\times\cr
&\Delta(J_{n-1,0}^\prime j J_n) \Delta(J_{n-1,1} j
J_n)(2J_{n-1,1}+1)\{J_{n-2,0}jJ_{n-1,1}/jJ_nJ ^\prime \}
\end{align}
Several terms in (49) are rational functions. Using the form (7) of
6\textit{j}-symbols, the following product
\begin{align}
        \Delta(J_{n-2,0}jJ_{n-1,1})\Delta(J_{n-1,0}^\prime j J_n) \Delta(J_{n-1,1} j J_n)\{J_{n-2,0}jJ_{n-1,1}/jJ_nJ ^\prime \}
\end{align}
 becomes equal to a rational function multiplying
\begin{align}
        &\Delta(J_{n-1,0}^\prime j J_n) \Delta(J_{n-1,1} j J_n)^2
\Delta(J_{n-2,0} j J_{n-1,1})^2\times \cr
&\Delta(j j J ^\prime )
\Delta(J_{n-2,0} J^\prime J_n)
\end{align}
Thus, (49) may be expressed as a rational function multiplied by
\begin{align}
 &N(J_{n-1,0}^\prime J_n)\Pi N(J_{r0}J_{r+1,0})\Pi
 N(J_{r0}^\prime J_{r+1,0}^\prime)\Pi[(2J_{r0}+1)(2J_{r0}^\prime+1)]^{1/2}\cr
&[(2J_{n-1,0}^\prime+1)(2J ^\prime+1)]^{1/2} \Pi \Delta
(J_{r0}^\prime jJ_{r+1,0}^\prime ) \Pi \Delta (J_{r1}jJ_{r+1,1})\cr
&\Delta(J_{n-1,0}^\prime j J_n) \Delta(j j J
^\prime)\Delta(J_{n-2,0} J^\prime J_n)
\end{align}
The summation in (46) is over $J_{n-1,1}$  which does not appear in
(52) and hence, it appears only in the rational function multiplying
it. Therefore, the summation over it in (46) can be simply carried
out yielding another rational function multiplying the expression
(52).

Thus, the $n-2\rightarrow n$
c.f.p.$[j^{n-2}(a_{n-2,0}J_{n-2,0})j^2(J ^\prime)J_n|\}j^n a J_n]$
is the square root of a rational function and its square is a
rational function. According to (45), the expectation value of a
state with these c.f.p. of a two-body interaction is a linear
combination of two-body energies with rational coefficients.

The $n-2\rightarrow n$ c.f.p. obtained above, may be used to derive
expressions of non-diagonal matrix elements of a two-body
interaction. Instead of (45), such non-diagonal elements are defined
by
\begin{align}
&<j^n a J|\Delta V_{ik}|j^n a^\prime J >=\cr &[n(n-1)/2]
\Sigma[j^{n-2}(bJ_1)j^2(J^\prime )J|\}j^naJ]\times \cr
                       &[j^{n-2}(bJ_1)j^2(J^\prime )J|\}j^na^\prime J]<j^2J^\prime |Vj^2J^\prime >
\end{align}
Only non-vanishing terms are written on the r.h.s. of (53). The
expression (52) is taken to be the non-rational factor in the c.f.p.
of the state \textit{a}. The c.f.p. of the state $a^\prime $ is the
same expression in which $J_{r0}^\prime $ is replaced by $J_{r0}"$.
All angular momenta $J_{r0}$ which define the$|j^{n-2}bJ_{n-2,0}>$,
as well as $J^\prime$, appear in the same expressions in both c.f.p.
and hence, contribute in (53) only rational functions. Each term in
the summation in (53) is thus equal to a rational function
multiplying the expression
\begin{align}
      &N(J_{n-1,0}^\prime J_n^\prime)N(J_{n-1,0}"J_n) \Pi N(J_{r0}^\prime J_{r+1,0}^\prime)\Pi
      N(J_{r0}"J_{r+1,0}")\times \cr
&\Pi[(2J_{r0}^\prime +1)(2J_{r0}"+1)]^{1/2}\Pi\Delta (J_{r0}^\prime
jJ_{r+1,0}^\prime ) \Pi \Delta (J_{r0}"jJ_{r+1,0}")\times \cr
&\Delta(J_{n-1,0}^\prime j J_n) \Delta (J_{n-1,0}"j J_n)V(J^\prime)
\end{align}
The factor of $V(J^\prime)$in (54) is a common factor of all terms
in the summation in (53). Hence, the non-diagonal matrix element
(53) is a linear combination of two-body energies whose coefficients
are rational functions multiplied by a common, $J^\prime$
independent factor. The square of this factor as seen in (54) is a
rational function of the angular momenta involved.

\section{The seniority scheme}
\par
The seniority quantum number \textit{v} measures in some sense the
amount of pairing of particle pairs in \textit{J}=0 states. In the
seniority scheme, states are eigenstates of the pairing interaction
defined by
\begin{align}
<j^2JM\mid q_{12}\mid j^2JM>=\delta_{J0}\delta_{M0}(2j+1)
\end{align}
The eigenvalues of this interaction in the $j^n$ configuration are
equal to
\begin{align}
(n-v)(2j+3-n-v)/2
\end{align}
Thus, loosely speaking, \textit{v} is the number of unpaired
particles. There is a very large class of interactions, not just the
pairing interaction, which are diagonal in the seniority scheme.
Also for general interactions, the seniority scheme is a convenient
one to use. Here, this scheme will be discussed only in relation to
the theme of the paper. The aim is to see which of the states of
this scheme have diagonal matrix elements which are linear
combinations of two-body energies with rational  coefficients.

According to (56), states with seniority \textit{v} in the $j^v$
configuration have no \textit{J}=0 pairing. States with the same
seniority in higher configurations are obtained by starting from
those states, adding $(n-v)$/2 pairs with \textit{J}=0 and
antisymmetrizing. Adding one particle to a state with seniority
\textit{v} and antisymmetrizing, the state obtained has seniority
\textit{v}+1 or \textit{v}-1. It may also be a linear combination of
states with both seniorities. The state with no \textit{j}-particles
has seniority \textit{v}=0 whereas the one particle state, with
\textit{J=j}, has \textit{v}=1. States with two particles have
states with \textit{J}=0, \textit{v}=0 or $J>$0, \textit{even} and
seniority \textit{v}=2.

Consider a state in the $j^n$ configuration, $n<2j$ \textit{even}
with \textit{J}=0 and \textit{v}=0. Adding one particle to this
principal parent, a state with \textit{J}=\textit{j} and
\textit{v}=1 is obtained. A state, with $n<2j$+1 odd, with $J=j$,
\textit{v}=1, is a principal parent of a \textit{J}=0, \textit{v}=0
state. Hence, such states have the property considered in this
paper. The expectation values of any two-body interaction of states
with lowest seniorities, \textit{v}=0 or \textit{v}=1, are given
explicitly by (30.1) ref.[2] or (20.13) ref.[5] as
\begin{align}
an(n-1)/2+[n/2]b
\end{align}
where [\textit{n}] is the largest integer not exceeding \textit{n}.
In (57), \textit{a} and \textit{b} are linear combinations of the
\textit{V}(\textit{J}) which appear only in two linear combinations

 $ V_0=V(J=0)\quad $ and $\quad \bar{V}_2=\Sigma (2J+1)V(J)/ \Sigma(2J+1)\, \qquad J>0,
  even$

The coefficients \textit{a} and \textit{b} are given by these
combinations as
\begin{align}
  a=\{(2j+2)\bar{V}_2-{V}_0 \}/(2j+1)\quad b=(2j+2)(V_0-{\bar{V}}_2)/(2j+1)
\end{align}
\par
In the case of an interaction which is diagonal in the seniority
scheme, the expressions (57) are the eigenvalues of these states.
Unlike the cases of $j<$7/2, there are no closed formulae for
diagonal matrix elements of states with higher values of \textit{v}.
Similar expressions exist only for the average values of diagonal
matrix elements of all states with given \textit{v} in the $j^n$
configuration $v=n, n-2,\ldots$. It will be shown, however, that
states of the seniority scheme may be constructed from principal
parents, antisymmetrizations and orthogonalization if necessary. As
shown in the preceding section, this guarantees that they have the
property considered here. Diagonal matrix elements of a two-body
interaction of states in the seniority scheme, constructed as
explained above, are linear combinations of two-body energies with
rational coefficients.

In the $j^2$ configuration, the \textit{J}=0 state has seniority
\textit{v}=0 as mentioned above. The other antisymmetric states have
even values of $J>$0 and seniority \textit{v}=2. As mentioned above,
adding a \textit{j}-particle to a principal parent with
\textit{J}=0, \textit{v}=0, the state $j^n$ $J=j$ with seniority
\textit{v}=1 is obtained. Since \textit{n-v} is always an even
integer, adding a \textit{j}-particle to a state with seniority
\textit{v} in the $j^n$ configuration (and antisymmetrizing) yields
a state of the $j^{n+1}$ configuration with seniority \textit{v}+1
or \textit{v}-1 or both. The added particle either completes a
\textit{J}=0 pair in which case the seniority is reduced or destroys
a \textit{J}=0 pair which raises the seniority.

If a \textit{j}-particle is added to a principal parent with
$J_0>0$, \textit{v}=2, the state obtained has \textit{v}=3 if $J\neq
j$. If a state with $J=j$ is constructed, it is a linear combination
of states with seniorities \textit{v}=1 and \textit{v}=3, if there
is a state with $J=j$, \textit{v}=3 in the $j^3$ configuration. Such
a linear combination will have a non-vanishing c.f.p.
$[j^2(v=0,J=0)J=j|\}j^3J=j]$. The $J=j$, \textit{v}=1 state obtained
earlier, may be projected out by adding it with a coefficient that
was defined by (13). The remaining state will be a pure $J=j$,
\textit{v}=3 state, orthogonal to the $J=j$, \textit{v}=1 state.
This procedure was described in detail in the preceding section. If
there are several $J=j$ states in the
 $j^3$ configuration, they may be constructed from several principal parents.
 They should then be orthogonalized to the $J=j$, \textit{v}=1 state and among themselves.

This procedure may be carried out in any $j^n$ configuration. If
\textit{n} is even, the steps described above may be taken. States
with seniorities \textit{v}=4 to which a \textit{j}-particle is
added, may be used to construct \textit{v}=5 states or linear
combinations of \textit{v}=3 and \textit{v}=5 states. In that case,
the components with \textit{v}=3 may be projected out by using the
\textit{v}=3 states constructed in the preceding step. At the end,
states with seniority $v=n$ may be used to construct states with
seniorities $v=n$+1 in the $j^{n+1}$ configuration.

If \textit{n} is odd, the state with $J=j$, \textit{v}=1 may be used
to construct a state with \textit{J}=0 which has seniority
\textit{v}=0, there are no states with \textit{J}=0 and
\textit{v}=2. States with seniority \textit{v}=2 may be constructed
from the same $J=j$, \textit{v}=1 state by coupling an additional
\textit{j}-particle to obtain a state with \qquad \qquad \qquad
\qquad $2j
>J
>0$, \textit{even}. From states with \textit{v}=3, states with \textit{v}=4 may be obtained. If
a linear combination of \textit{v}=2 and \textit{v}=4 is obtained,
the \textit{v}=2 components may be projected out by using
\textit{v}=2 states obtained earlier. This procedure may be further
carried out until all states in the $j^{n+1}$ configuration are
constructed. They all have well defined seniorities and their
diagonal matrix elements have the property described in this paper.

An example of the procedure described above is offered in ref.[1].
The authors calculated the eigenvalues of \textit{J}=2 and
\textit{J}=4 states with seniority \textit{v}=4 in the (7/2)$^4$
configuration. They found that the coefficients of two-body energies
in both cases are rational numbers. In that configuration, there are
also \textit{J}=2 and \textit{J}=4 states with seniority
\textit{v}=2. The latter may be obtained from the state in where a
7/2 particle is coupled to the principal parent with \textit{J}=7/2
in the (7/2)$^3$ configuration. There is only one such state in that
configuration and hence, it has seniority \textit{v}=1. The
\textit{J}=2 and \textit{J}=4 states obtained from it have seniority
\textit{v}=2 since the only state with \textit{v}=0 has
\textit{J}=0. The other \textit{J}=2,4 states must be obtained from
another principal parent. In that one, a 7/2 particle should be
coupled with any \textit{v}=3 state in the \qquad \qquad(7/2)$^3$
configuration. From the resulting states, the \textit{v}=2
components should be projected out as described above. Hence, the
\textit{v}=4 states are examples of states which have the property
considered in the present paper.

A more complicated situation occurs in the (9/2)$^4$ configuration
considered in ref.[1]. In that case there are 3 states with
\textit{J}=4 and 3 states with \textit{J}=6. One of the \textit{J}=4
(and \textit{J}=6) states has seniority \textit{v}=2 and may be
constructed from a state where a 9/2 particle is coupled to the
\textit{J}=9/2, \textit{v}=1 state. One of the remaining
\textit{J}=4 states may be constructed from a principal parent with
\textit{J}=9/2, \textit{v}=3 coupled to a 9/2 particle. The
resulting state,$|a\!>$, is a linear combination of \textit{J}=4,
\textit{v}=4 and \textit{J}=4, \textit{v}=2 states. Another
independent \textit{J}=4 state,$|b\!>$, may be constructed from
another principal parent with \textit{v}=3 and is also a combination
of \textit{v}=4 and \textit{v}=2 states. The \textit{J}=4,
\textit{v}=4 states which are discussed in ref.[1] may be obtained
as follows. First, the state $|b>$ is orthogonalized to state $|a>$
as indicated by vanishing of the c.f.p.
$[(9/2)^3(J_0=9/2,v=3)(9/2)J=4|\}(9/2)^4bJ=4]$ of the orthogonalized
$|b>$ state. Then, from both states the \textit{J}=4, \textit{v}=2
components are projected out leaving them as pure \textit{v}=4
states. In each resulting state the c.f.p. corresponding to that of
the principal parent of the \textit{v}=2 state,
$[(9/2)3(J_0=9/2,v=1)(9/2)J=4|\}(9/2)^4J=4]$, vanishes. It follows
that these \textit{v}=4 states have the property considered in this
paper. The same arguments apply also to the \textit{J}=6,
\textit{v}=4 states. Their diagonal matrix elements are linear
combinations of two-body energies with rational coefficients.
\section{Summary}
In the present paper, matrix elements of a two-body interaction
between states of $j^n$ configurations were considered. The two-body
interaction is taken to be a perturbation and hence, only the
sub-matrix of the Hamiltonian, defined by states of the  $j^n$
configuration, is considered. The matrix elements are functions of
two-body energies $V(J)=<j^2JM|V|j^2JM>$. There are states whose
expectation values (diagonal matrix elements) are linear
combinations of the $V(J)$ with positive rational coefficients. In
this paper, it was shown that this is always the case for states
which are the only ones with given \textit{J} in the  $j^n$
configuration considered. Such states are eigenstates of the
sub-matrix considered.

When there are several states with the same value of \textit{J}, the
situation is more complicated. It was shown above that it is
possible to construct in such cases, a complete basis of states
which have this simple feature. The diagonal matrix elements of
these states are linear combinations of the \textit{V(J)} whose
coefficients are positive rational numbers. The construction of
these states was carried out by using coefficients of fractional
parentage (c.f.p.). These are obtained by starting from states of
the $j^3$ configuration which serve as principal parents. This
procedure is repeated by adding more particles until the $j^n$
configuration is reached. In every step, orthogonalization (and
normalization) of the states thus obtained, is performed. It was
shown that the squares of the c.f.p. are rational functions of the
angular momenta used in the construction and hence, are independent
of the particular two-body interaction.

It was shown that states of the seniority scheme form bases of this
kind. If the interaction is diagonal in the seniority scheme,
non-diagonal matrix elements between states with different
seniorities vanish. This, however, is not true for the most general
two-body interaction. In any case, there are usually non-vanishing
non-diagonal matrix elements between states with the same seniority
\textit{v} and the same spin \textit{J}. Irrespective of the nature
of the basis states for a given value of \textit{J}, they define a
sub-matrix of the Hamiltonian. This sub-matrix should be
diagonalized in order to find the eigenstates and corresponding
eigenvalues.

The eigenstates are linear combinations of the basis states but the
coefficients strongly depend on the eigenvalues. The latter are
functions of the \textit{V(J)} but are usually far from linear
functions. The eigenvalues are the roots of the secular equation
whose degree is limited only be the order of the sub-matrix. The
eigenstates may be expanded in terms of c.f.p. but unlike the basis
states constructed in this paper, their c.f.p. are functions of the
\textit{V(J)}.

It is convenient to use the basis states to set up the sub-matrix to
be diagonalized. In addition to the rather simple structure of
diagonal elements, also the non-diagonal elements have a simple
structure. Each is a linear combination of the $V(J)$ with rational
coefficients multiplied by a factor which is the square root of a
rational function

From the discussion above, a simple conclusion follows that states
with the rational property are the exception rather than the rule.
In any given $j^n$~configuration, states which are the only ones
with a given value of \textit{J} are rather rare. They may be found
among the states with the highest values of \textit{J} (certainly
the state $J=nj-n(n-1)/2)$ as well as among those with the lowest
\textit{J} values.

Still, there may be special cases like those mentioned in ref.[1].
In the case of $j=7/2$, the non-diagonal matrix element of any two
body interaction between the \textit{J}=2 states with \textit{v}=2
and \textit{v}=4 vanishes. The same is true also for the
\textit{v}=2 and \textit{v}=4 states with \textit{J}=4. This follows
from general properties of the seniority scheme. More surprising is
the fact that in the $(9/2)^4$ configuration, there is a
\textit{v}=4, \textit{J}=4 state whose matrix element of any
two-body interaction, vanishes not only with the \textit{v}=2,
\textit{J}=4 state but also with the orthogonal \textit{v}=4,
\textit{J}=4 state. Its eigenvalue is a linear combination of
two-body interactions with positive rational coefficients as given
in ref.[1]. The same situation occurs also for the corresponding
\textit{J}=6 states.


\begin{thebibliography}{99}
\bibitem{1}
L.Zamick and P.Van Isacker, Phys. Rev. \textbf{78}, 044327 (2008).
\bibitem{2}A.de Shalit and I.Talmi, Nuclear Shell Theory,
     Academic Press (1963); Reprinted by Dover
     Publications (2003).
\bibitem{3}R.F.Bacher and S.Goudsmit, Phys. Rev. \textbf{46} (1934) 948.
\bibitem{4}G.Racah, Phys. Rev. \textbf{63} (1943) 367.
\bibitem{5}I.Talmi, Simple Models of Complex Nuclei, Harwood
    Academic Publishers (1993).
\bibitem{6}G.Racah, Phys. Rev. \textbf{62} (1942) 438.
\end{thebibliography}
\end{document}